# Supervised laser-speckle image sampling of skin tissue to detect very early stage of diabetes by its effects on skin subcellular properties

**Ahmet Orun[1], Luke Vella Critien[1], Jennifer Carter[2] and Martin Stacey[1]**

[1]De Montfort University, Faculty of Technology, Leicester, LE1 9BH, UK
aorun@dmu.ac.uk, luke.vella.critien@ensquad.com, mstacey@dmu.ac.uk
[2]J.Carter@hud.ac.uk

**Abstract**

This paper investigates the effectiveness of an expert system based on *K*-nearest neighbours algorithm for laser speckle image sampling applied to the early detection of diabetes. With the latest developments in artificial intelligent guided laser speckle imaging technologies, it may be possible to optimise laser parameters, such as wavelength, energy level and image texture measures in association with a suitable AI technique to interact effectively with the subcellular properties of a skin tissue to detect early signs of diabetes. The new approach is potentially more effective than the classical skin glucose level observation because of its optimised combination of laser physics and AI techniques, and additionally, it allows non-expert individuals to perform more frequent skin tissue tests for an early detection of diabetes.

## 1 Introduction

An expert system has been developed for instant automated laser speckle image sampling that can detect the changes in micro-level subcellular skin properties like cell nucleus, membrane, etc. The aim of the work is to develop a novel method for early diagnosis of diabetes through an automated process which quantizes the sub-cellular parameters via texture measures[4] applicable to laser speckle imagery and then classifies the laser speckle image into three classes: Normal, Type 1 Diabetes and Type 2 Diabetes, to detect early signs of disease. At the first stage, the proposed system does not target a maximum classification accuracy, it rather aims to demonstrate an optimisation by which optimum system parameters and options are automatically selected, which would then be ready to process any test data from any sources to maximize its classification accuracy. The system would also allow inexperienced users and non-clinicians to make frequent observations on easily accessible regions of skin, such as patients' forearms, to detect subtle changes in skin subcellular properties.

### 1.1 Laser speckle imaging of skin

Laser light with specific wavelength characteristics can be reflected or scattered from a surface which presents a particular intensity distribution that consists of randomly generated alternately dark and bright spots of variable shapes[8]– a speckle pattern. The physical property of laser light (e.g. wavelength and energy) determine the nature of its interaction with a textural structure of skin cells[1,2] whose speckle effects provide critical information about the specific characteristic of skin subcellular properties via laser speckle images. Such coded images are then processed by an artificial intelligence method (e.g. *k-NN* algorithm, Bayesian networks or neural nets) to unveil such information that refers to early signs of diabetes.

The laser speckle phenomenon is described by a particular intensity distribution of a light reflected and scattered from a rough surface (such as human skin). This speckle occurrence is a stochastic effect and can only be described statistically[8] by a formula (1).

$$C = \sigma / \langle I \rangle \qquad (1)$$

where C is the contrast of the speckle, $\sigma$ is the standard deviation of the spatial intensity of the speckle pattern and $\langle I \rangle$ is the main intensity of the speckle pattern. When the value of C is equal to 1 then the speckle pattern has the maximum contrast which means it is fully developed. The value of contrast C can be in the range of 0 to 1 depending on the skin surface characteristics.

## 1.2 Effects of diabetes on skin

Diabetes mellitus is a medical term for what is commonly referred to as diabetes. The term "mellitus" is of Latin origin meaning honey or sweet – referring to the surplus sugars which are detected in blood or urine samples of diabetic patients.[9] With records dating back to the year 400BC, diabetes is considered one of the oldest diseases to be discovered. Diabetes mellitus is nowadays considered one of the top four most dangerous systemic disease states.[5] Despite a variety of screening methods used over the years to detect and monitor this disease, some patients still remain undiagnosed, especially in the early disease stages. According to Ghosh et al[10] "Skin manifestations in diabetes are very common and a frequent subject of study, though often neglected in clinical grounds". One of the major reasons giving rise to skin problems in diabetic patients, is due to the skin turning dry. The body attempts to discard excessive blood glucose by turning water into urine, hence reducing fluid levels within the body and resulting in the skin turning dry and damaged. Grandinetti and Tomecki[11] found that "between 30% to 50% of diabetic patients will develop a skin condition" and ''skin is often a window to systemic disease''. Research indicates that diabetic patients suffer from an array of skin disorders. Literature suggests that amongst the most common diabetic skin conditions detected one finds: diabetic dermopathy, bacterial and fungal infections, eruptive xanthomatosis, and necrobiosis lipoidica diabeticorum. Goyal et al[12] and Foss et al[13] clearly state that there is a strong correlation between skin conditions and diabetes. The earlier study conducted by Goyal et al[12] indicated that skin manifestations should prompt healthcare professionals to examine a patient's diabetic status. This could be the key to treatment and prevention as indicated by Foss et al[13] "poor metabolic control of diabetes increases patient's susceptibility to skin infections" amongst other complications.

Image based skin monitoring techniques are very common to diagnose skin diseases and conditions[3] particularly real-time observation of micro blood perfusion. But such methods are qualitative and rather for real-time observation purposes. Mahe et al[14] used the laser speckle phenomenon to generate a perfusion map of the tissue with different pseudo colours for visual observation. Even though a major advantage of LSI instrumentation is its inexpensive set up, its specifically developed software utilities are at high cost, which makes such systems not cost effective. Richards et al[15] and Jakovels et al[16] present two studies where LSI techniques are used with similar low cost instrumentation.

## 2 The laser speckle imaging classification technique

The trials of the laser speckle imaging technique[2] for diagnosing diabetes employed laser light at $\lambda$=650nm wavelength which is the most suitable for skin analysis for sub-skin structure.[6,7] At system operational level, a continuous loop is used for iterative image processing and analysis which is comprises five stages namely: Image Pre-Processing, Pilot Sample Detection, Testing Sample Detection, Texture Analysis (Feature Extraction) and Classification. The use of histogram thresholding and k-means clustering were fundamental to identifying the energy bands within the research images. The template matching technique was examined and explored to extract samples within the test images. A set of textural features[4] were then used to create the data sets for classification. The classification stage made use of the k-nearest neighbour algorithm with a varying distance option. Throughout the study these stages were optimized to yield an efficient application that could automatically select the best classification[10] data set.

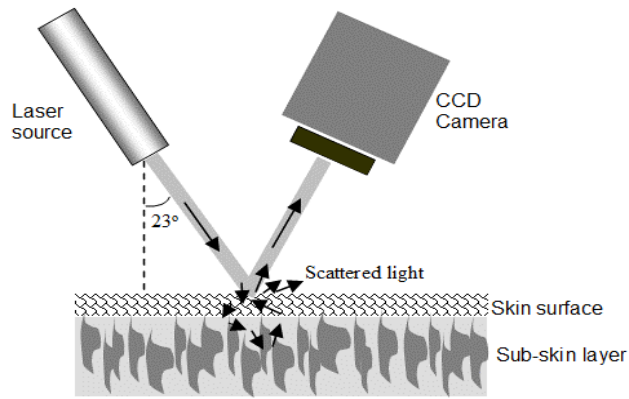

Figure 1. LSI basic instrumentation (Orun et al.[2])

## 2.2 Hardware used in the experiments

The series of speckle images were taken from the (normal and diabetic) subjects' forearm regions in a non-invasive way, using a commercial CCD image camera and a laser source. The images were at *3840 x 2880* pixel resolution. The laser source used was at an approximately 23° angle to the skin surface normal (Figure 1). The laser light was from a collimated red laser diode at 650 nm wavelength. The power was quite low (1 mW), much less than maximum permissible exposure which is 2000 W m² for human skin (for 400–700 nm wavelength range) for a long term (5–10 h) exposure. The laser light used in our experiments was operated only for short-term illuminations (approximately 2-3 seconds) over an approximately 10 mm diameter area on forearm skin of each participant to generate a speckle effect on the skin surface.

## 2.3 Texture analysis

The main application loop involves a texture analysis stage. This is because a laser speckle image generated by laser-skin interaction has a textural pattern which is specific to skin's cellular structure, subcellular components, etc. Texture analysis was vital to the whole process as the textural features it produces are essential for the process of classification. This study made use of different texture characteristics derived from works by Phillips[4] and Haralick et al[19] which both contributed to a number of texture measures which have been tried and tested in different experiments on textural analysis. The first set of textural measures used within this work was derived from Phillips.[4] The mathematical expressions described in Equations 2 – 6 were programmed and their results were later used in the data set generation.

$$Variance_{russ} = \sqrt{\sum_{(Russ)}(centerpixel - neighbour)^2} \qquad (2)$$

$$Variance_{levine} = \frac{1}{area} \sum_{(Levine)}(centerpixel - mean)^2 \qquad (3)$$

$$\sigma = \sqrt{Variance_{Levine}} \quad (Sigma) \qquad (4)$$

$$Skewness = \frac{1}{\sigma^3} \frac{1}{area} \sum_{(Skewness)}(centerpixel - mean)^3 \qquad (5)$$

$$Standard\ Deviation = \sqrt{\frac{\sum(x - a')^2}{n}} \qquad (6)$$

The second set of textural measures implemented in this study was derived from Haralick et al.[19] In the work, fourteen 'Haralick features' were extracted via a Gray Level Co-occurrence Matrix (*GLCM*). The contrast, correlation, energy, homogeneity and entropy features are currently considered to be the most significant 'Haralick features' with respect to texture analysis[19].

## 2.4 Pre-processing and pilot sample detection

The pre-processing stage involved four major tasks: pilot image loading, grayscale conversion, histogram creation and detection of segmentation templates within the speckle image. This preparatory stage generated a number of intermediate products which were required in successive stages. Firstly the system user was required to select the *pilot image* – the main image from which samples were later extracted. After pilot image was selected, it was translated into grayscale. This grayscale transformation was a necessary step performed on all the images used within the application to drastically reduce the required processing power (which is the major resource bottleneck for using this method for diagnosis). The grayscale conversion was followed by the generation of a histogram. Image histograms present a pictorial representation of the tonal distributions within an image, as illustrated in Figure 2. Histogram generation was required to determine the number of energy bands within the pilot image. The histogram in Figure 2 clearly displays the selected pilot image as having four major peaks and three valleys. The spikes are the points of a large concentration of a particular colour. The peaks are used to determine where the regions reside within the image.

The final task of the pre-processing stage included the generation of two segmented images using two different image segmentation techniques – threshold segmentation and K-means clustering. The number of valleys within the histogram was determined by eye by the user. The valleys value was important in the generation of the two segmentation templates, which are used to divide the pilot image into regions that are used to determine the locations from which the sample areas used for classification are taken. The implemented histogram thresholding method involved multiple thresholds that were automatically derived from the valley points in the histogram. Multiple threshold histogram segmentation is more complex than the commonly used binary thresholding. The second segmentation method implemented was K-means clustering, which used the same histogram information to generate the segmentation template. This unsupervised method was executed for three successive times in order to reduce the possibility of local minima. On completion of these two image segmentation templates, the application displayed the results beneath the original pilot image and the generated histogram as shown in Figure 2. Various similarity metrics can be applied to determine the similarity between the template (segment) and the area (in whole image) being tested. Hisham et al[20] compared the Sum of Squared Differences (SSD) and Normalised Cross Correlation (NCC) similarity metrics. They concluded that although both SSD and NCC gave similar results, NCC was more robust. The NCC metric is expressed using the equation (7) where $f$ is the image, $\bar{t}$ is the mean of the template and $\bar{f}_{u,v}$ is the mean of $f(x,y)$ in the region under the template.

$$y(u,v) = \frac{\sum_{x,y}[f(x,y)-\bar{f}_{u,v}][t(x-u,y-v)-\bar{t}]}{\left\{\sum_{x,y}[f(x,y)-\bar{f}_{u,v}]^2 \sum_{x,y}[t(x-u,y-v)-\bar{t}]^2\right\}^{0.5}} \qquad (7)$$

Based on this conclusion, the application made use of the NCC similarity metric. NCC returned a value with a maximum range of 1 for each comparison performed. The sample with the best NCC value was subsequently chosen as the best matching sample within the test image.

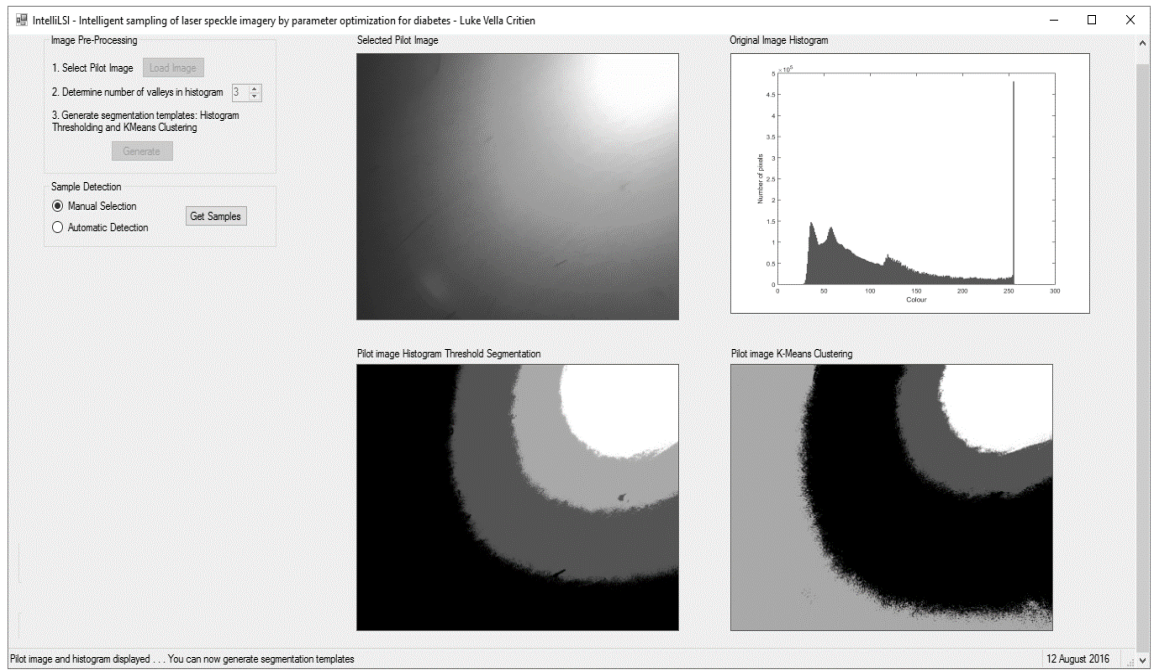

Figure 2. Application status after the pre-processing and sample detection stages

## 2.5 Image classification method

The image classification process, which used the k-nearest neighbours algorithm, could distinguish between normal and diabetic skin characteristics as each of them is specified by subcellular changes such as deformed cell nucleus, changes in cellular substance, etc. K-nearest neighbours *(k-NN)* is one of the simplest classification algorithms with a vast range of applications. The results of this classifier were calculated on the distance of the object from its neighbours (a metric of how much it differs from the most similar reference objects), calculated from the image texture measures discussed in section 2.3. Various research studies in the past have used this technique to classify different medical cases such as lymph node metastasis in gastric cancer[18], melanoma recognition[18] and many others.

## 2.6 The operation of the system

After the guided image sampling stage in which optimum sampling style was learned by the system, it ran iterative loops where each one calculated the statistical texture measures of a speckle image (as shown in Figure 1), then performed classification to obtain the best classification accuracy result. Each time the application loop was executed, all the steps discussed in this section were performed. The number of tests carried out by the application was dependent on the number of test images that were selected by the user. For each test performed by the classification algorithm, the steps below were followed:

- For each sample detected through the NCC template matching stage, seventeen features were calculated. Philip's first four features (represented by equations 2-5)[4] were calculated for two separate operational windows – a *3x3* and *5x5* window respectively. These features therefore accounted for eight of the seventeen features. The sample mean, median and standard deviation were another three features. The remaining six features, included the image intensity and Haralick's five textural features (represented by equations 7 -11)[19]

- The total number of pilot samples and their respective features (mentioned in the previous step) were equally divided into training and testing data sets. These data set files included an equal number of samples from each image type (normal, diabetes type 1 and diabetes type 2) that were randomly selected. In this step two training and two testing data sets were generated. Both files had the same number of rows but possessed varying features. The training/test files contained Philip's eight features together with the mean, median and standard deviation while the *GLCM* training/test files included all the seventeen features.

- A further two files – training class and test class – were generated, where each one had a single column which included a value between 1 and 3 for each row in the testing and training data sets. These two files were needed by the classification algorithm to enable distinction among the different image types.

The application loop required a minimum of eight images (four from each image type) to be selected in order to successfully perform the classification process. For each of the three consecutive runs, the number of tests executed depended on the number of test images selected. If the number of selected test images was eight (the minimum allowed) the tests in each run would only be two. The first test would use the training data set as the training input and the testing data set as testing input. The second test would have the training and testing data set switching position in the *KNN* classifier training and testing parameters. For each extra test image (apart from the eight) selected, another two tests were executed, where the sample images and their features randomized from scratch. This decision was taken to allow for a new training and testing data set to be generated and hence creating different possibilities to be tested in the process. The complete configuration of system operation is shown in Figure 3.

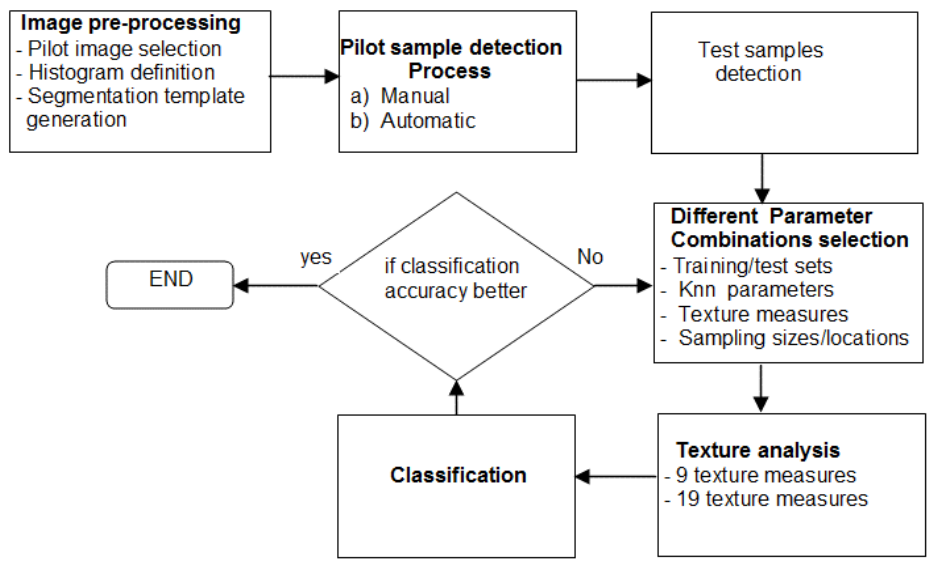

Figure 3. The flowchart of system operation with an optimization loop for different parameter combinations

## 3 Results and discussion

The classification accuracy between normal, diabetes A and diabetes B classes was the most important success measure and applied onto the different stages of the system process. This included the sample sizes used for speckle (pattern) image sampling and their consequent classification results as shown in Figure 4. According to results, a correlation exists between the sample size and classification accuracy for the sampling process that includes all classes together (normal, diabetes Type 1 and Type 2). Figure 5 also exhibits the classification accuracy results by the selection of different data set attributes (texture measures) for 1) Philip's 11 texture measures[4], and 2) Haralick's 17 texture measures.[19] In Figure 4, 5 and 6, diabetes Type 1 is cited as "diab" and diabetes Type 2 is cited as "diabB".

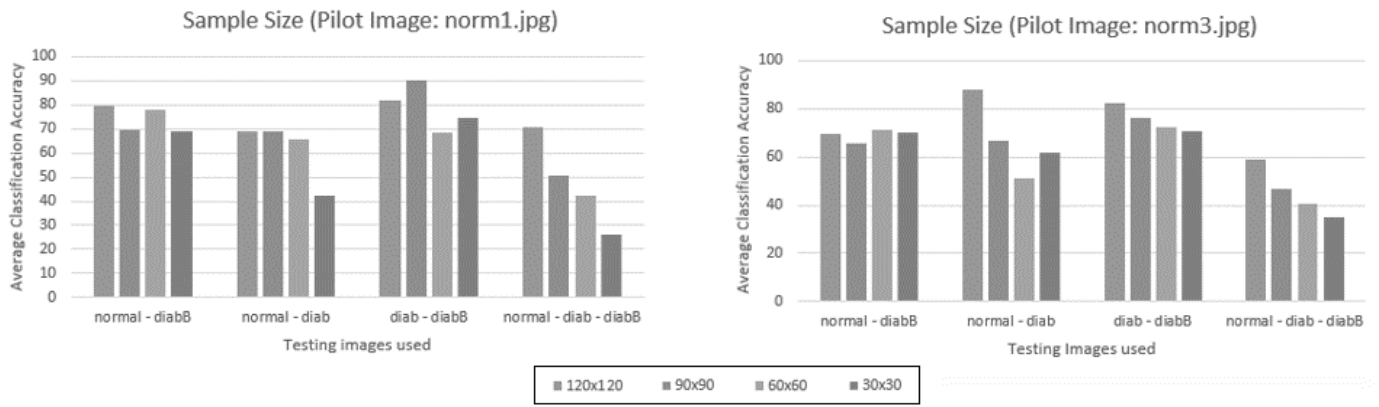

Figure 4. The sample sizes used for speckle (pattern) image sampling and their consequent classification results. The correlation between the sample size and classification for all classes is exhibited only in the last section of the graph. (image sampling sizes *120x120, 90x90, 60x60 and 30x30* are in the same order with the individual graph bars 1- 4 for all categories respectively)

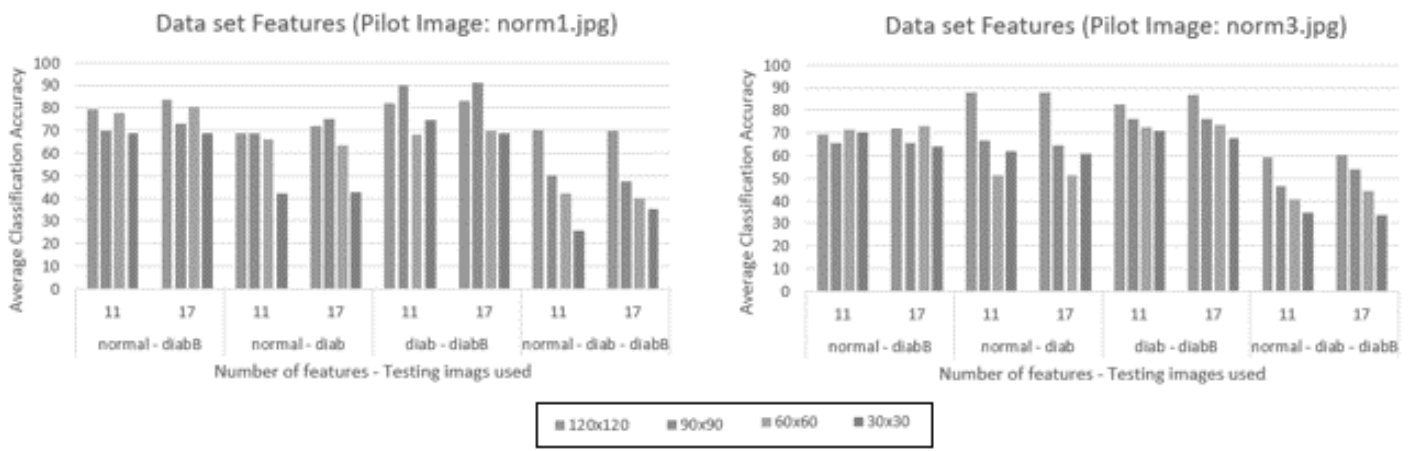

Figure 5. Classification accuracy results by the selection of different data set attributes (texture measures) by two categories 1) Philip's 11 measures[4] and 2) Haralick's 17 measures[19]. (image sampling sizes *120x120, 90x90, 60x60 and 30x30* are in the same order with the individual graph bars 1- 4 for all categories respectively)

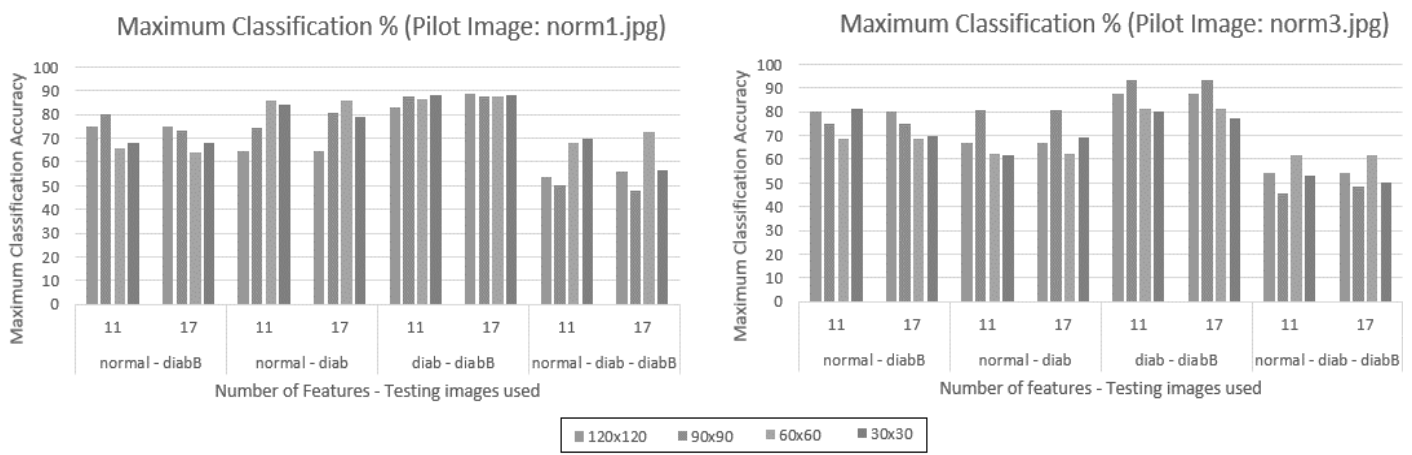

Figure 6. The maximum Classification accuracy results obtained by the selection of different data set attributes (texture measures) by two categories ( 11 measures and 17 measures). (image sampling sizes *120x120, 90x90, 60x60 and 30x30* are in the same order with the individual graph bars 1- 4 for all categories respectively)

The histograms shown in figure 6 outline the maximum value obtained through tests, where the average classification accuracy generated through the consecutive tests was considered. From the graphs in Figure 6, it was evident that the best maximum classification was obtained (90%) when the application made use of the pairs of "normal - diabetes type 1" and "diabetes type 1- diabetes type 2" images as test images. One can infer that these tests had the fewest test images and hence there was less variation between the samples tested leading to more accurate classification. The worst performing tests were recorded when the application considered all the images from the three different categories (normal - diabetes type 1 - diabetes type 2) as the testing images. Notwithstanding a low maximum classification accuracy within this category, accuracy values of above 60% were still recorded.

The generation of results was a lengthy process (sometimes involving days) requiring all the available processing resources. The study made use of a laptop with an Intel Core i7-4700MQ processor, with 16GB DDRIII memory and a 512GB solid-state hard disk under Microsoft Windows 10. The complexity of the algorithms within the loop, such as image raster scans and API calls, coupled to the fact that the pilot and testing images were kept in their original size, were amongst the main reasons behind such a lengthy process.

## 4 Conclusion

The novel approach adopted in this study has led to the development of an optimized application which automatically detects the best combination of training data set among the whole set, based on the related classification accuracy results obtained (e.g. about 90%). The adopted approaches included swapping the testing and training data sets, using data sets with eleven[4] or seventeen[19] texture features, as well as the adoption of different *k-NN* parameters.

The main aim behind this study was to automate and simultaneously optimise the LSI classification process for detecting the presence of diabetes non-invasively from skin properties. This innovative approach has been successfully implemented and tested through the integration of a laser speckle imaging technique, classification algorithm and Haralick's and Philips textural measures within the study. This approach has never been adopted nor tested for diabetes before. It has been clearly demonstrated that positive results can be obtained by automating and successively optimising the classification process of diabetic skin images generated through low-cost and non-invasive instrumentation.

Further research would investigate how to use more specific laser types (using more precisely optimized wavelength and intensity, or a combination of images produced with different wavelength and intensity) which would lead to more effective interaction with the specific subcellular properties (e.g. insulin content, DNAs, etc.) to unveil more and earlier symptoms of diabetes. This research could also be extended to investigate the diagnosis of other "skin related" systemic diseases (e.g. heart disease, kidney disorder, etc.).

Cost effectiveness is a major challenge for the proposed system. The system will comprise a digital camera and laser source, from which the images are directly acquired and processed by a software utility. However the anticipated hardware costs could be justified if the proposed system were used quickly by non-specialists for large–scale routine screening. The costs for the major functional hardware components would match the cost of the prototype at less than a thousand dollars; however comparable medical equipment with very similar hardware components (but only for sub-skin display purposes) is sold for £40,000. A more difficult problem might be carrying out the amount of computation required quickly enough for clinical use.